# Growth Mechanisms and Oxidation-Resistance of Gold-Coated Iron Nanoparticles

Sung-Jin Cho[1], Juan-Carlos Idrobo[2], Justin Olamit[2],
Kai Liu[2*], Nigel D. Browning[3, 4], and Susan M. Kauzlarich[1*]

[1]*Department of Chemistry, University of California, Davis, CA 95616*

[2]*Department of Physics, University of California, Davis, CA 95616*

[3]*Department of Chemical Engineering and Materials Science,*

*University of California, Davis, CA 95616*

[4]*National Center for Electron Microscopy, Lawrence Berkeley National Laboratory,
Berkeley, CA 94720*

Corresponding authors: kailiu@ucdavis; smkauzlarich@ucdavis.edu

**Abstract**

We report the chemical synthesis of Fe-core/Au-shell nanoparticles by a reverse micelle method, and the investigation of their growth mechanisms and oxidation-resistant characteristics. The core-shell structure and the presence of the Fe & Au phases have been confirmed by transmission electron microscopy, energy dispersive spectroscopy, X-ray diffraction, Mössbauer spectroscopy, and inductively coupled plasma techniques. Additionally, atomic-resolution Z-contrast imaging and electron energy loss spectroscopy (EELS) in a scanning transmission electron microscope (STEM) have been used to study details of the growth processes. The Au-shell grows by nucleating on the Fe-core surface before coalescing. The magnetic moments of such nanoparticles, in the loose powder form, decrease over time due to oxidation. The less than ideal oxidation-resistance of the Au shell may have been caused by the rough Au surfaces. However, in the pressed pellet form, electrical transport measurements show that the particles are fairly stable, as the resistance of the pellet does not change appreciably over time.

## Introduction

Magnetic nanoparticles are of interest for a wide variety of applications; for technology, as magnetic seals, printing, recording,[1-3] and for biology, as magnetic resonance imaging (MRI) agents,[4, 5] cell tagging and sorting.[6] In these areas of research, particle size, shape and surface properties are important. Great progress has been made in the production of a variety of magnetic nanoparticles.[7] For example, iron oxides such as $Fe_2O_3$ and $Fe_3O_4$ can be prepared as monodispersed surface derivatized nanoparticles;[8-12] Co and Fe can be prepared as nanoparticles as well as nanorods by solution methods.[13, 14] Of special interest are core/shell structured nanoparticles that could exhibit enhanced properties and new functionality, due to the close proximity of the two functionally-different components. Such structures not only are ideal for studying proximity effects, but are also suitable for structure stabilization as the shell layer protects the core from oxidation and corrosion. Additionally, the shell layer provides a platform for surface modification and functionalization, such as coupling the magnetic core through the shell onto organic or other surfaces, thus tuning their intrinsic magnetic properties and making them potentially bio-compatible.[15] There has been extensive work on magnetic core/shell nanoparticles where the magnetic core is $Fe_3O_4$ and the shell is a polymer which provides biocompatibility and long-term stability.[16] In the case of Fe as the core, there are examples of core/shell Fe/Au,[17, 18] Fe/Fe-oxide,[19] and Fe-oxide/Au.[20]

The synthesis of Fe/Au core/shell nanoparticles is of special interest for possible application towards sensors,[21] drug delivery and bio-detection technologies.[22] However, the structural integrity and chemical stability of such nanoparticles remain as the primary challenges for the synthesis and employment of this type of artificial nanostructures.

Fe/Au core/shell nanoparticles have been prepared by other groups and their properties explored. In previous work by Carpenter *et al*,[23] core/shell structured Fe/Au nanoparticles were synthesized by a reverse micelle method and characterized by x-ray diffraction (XRD), ultraviolet-visible spectroscopy (UV-vis), transmission electron microscopy (TEM) and magnetic measurements. The Au shell was expected to protect the Fe core and to provide for further organic functionalization. These nanoparticles had a size distribution of 5-15 nm diameter and average size about 10 nm. The x-ray diffraction pattern showed peaks assigned to Au and Fe, but no diffraction associated with oxide. The blocking temperature was reported to be 42 K. Other short reports have followed.[24-26] The oxidation of these core/shell nanoparticles was also studied by x-ray absorption spectroscopy (XAS) and the Fe core was shown to be extensively oxidized. The oxide was most similar to that of $\gamma$-$Fe_2O_3$.[27] It was proposed that the Fe nanoparticle may not be centered in the micelle, resulting in an asymmetric Au shell. An alternate explanation was that there may be grain boundaries in the Au shell that allow for diffusion of oxygen and oxidation of the metallic core. In the report by Kinoshita *et al*,[28] the same synthetic method was followed and the sample was characterized by the same methods, along with x-ray absorption near edge structure (XANES) and extended x-ray absorption fine structure (EXAFS). The XANES spectra were consistent with the core magnetic phase being primarily $Fe_3O_4$. Other studies have suggested that the Fe/Au nanoparticles may not be prepared via the reduction route using the reverse micelle method.[29] The key issues here are the chemical states of the core materials and whether the oxide forms during or after the synthesis process.

We have investigated the reverse micelle synthetic method further and have found that the structure of Fe/Au core/shell nanoparticles is not as simple as either of the previous reports indicated.[18] These nanoparticles showed higher blocking temperature (150 K) and Mössbauer results were best interpreted as Fe speciation of α-Fe, $Fe^{II}$, $Fe^{III}$, and FeAu alloy. In addition, we determined that these nanoparticles decomposed rather quickly to $Fe^{III}$.

In this work, we report a detailed study of the size and chemical state of the Fe core, the oxidation resistance characteristics of the Fe/Au core/shell nanoparticles and their origin due to the growth mechanisms. We have achieved Fe/Au nanoparticles with large enough Fe cores to exhibit ferromagnetism at room temperature. Using XRD, TEM, single particle Electron Energy Loss Spectroscopy (EELS), and other chemical analyses, we have confirmed that the initial core material is indeed α−Fe. Furthermore, using atomic-resolution Z-contrast imaging in a scanning transmission electron microscope (STEM), we have found that the Au shell grows by nucleating at selected sites on the Fe core surface before coalescing. The resultant Au shell has a rough surface, which could compromise its oxidation-resistance.

**Experimental Section**

All chemicals were purchased from Aldrich Chemical, Alfa-Aesar, or Fisher Scientific. They were used without further purification and nano pure water (Barnstead ultra pure water system D11931) was used throughout. All solvents were degassed by the freeze, pump, thaw method. Nanopure water was degassed by bubbling argon gas through the water for 2 hours.

*Synthesis of Nanoparticles*

Fe/Au nanoparticles were synthesized as previously reported.[17, 18] The reaction was carried out in a reverse micelle reaction under argon gas by utilizing Schlenk line anaerobic techniques. Cetyltrimethylammonium Bromide (CTAB) was used as the surfactant, octane as the oil phase, and 1-butanol as the co-surfactant. The water droplet size of the reverse micelle was controlled by the molar ratio of water to surfactant.

Iron nanoparticles were prepared by the reduction of $Fe^{2+}$ with $NaBH_4$. 0.18g (1.2mmol) of $FeSO_4$ was added to the inverse micelle solution and 0.09g (2.4mmol) of $NaBH_4$ in the reverse micelle solution was added via double-ended needle. The mixture was stirred at room temperature for 1 h. The dark powder was separated from the solvent with a magnet and washed with $CH_3OH$ twice and dried under vacuum. To create a gold shell on the Fe core, 0.27g (0.8mmol) of $HAuCl_4$ was prepared as a micelle solution and added to the solution of $FeSO_4$ and $NaBH_4$. 0.11g (2.9mmol) of $NaBH_4$ micelle solution was immediately added to the solution and it was left stirring at room temperature overnight. A dark precipitation was separated with a magnet and washed with $CH_3OH$ twice to remove any nonmagnetic particles and organic surfactant. The sample was dried in vacuum. The yield is 43 mg (19 weight % yield).

*Structure Analysis*

Powder X-ray diffraction measurements were made on a Scintag PAD-V diffractometer using Cu $K_\alpha$ radiation ($\lambda$ = 1.5418 Å) and Material Data Inc. (MDI) JADE6 software. The nanoparticles were packed on the glass sample holder in nitrogen gas-filled glove box, then sealed by clear tape to prevent air contact. The XRD patterns were collected between $30° < 2\theta < 90°$ with dwelling time of 2 seconds and step size of 0.02 ($2\theta$). XRD line widths were calculated using the MDI software to subtract background and $K_{\alpha 2}$ peak. The crystallite sizes of nanoparticles are determined by Scherrer equation: $L = (0.88\lambda)/(\beta \cos\theta)$, where $\lambda$ is the X-ray wavelength in nm, $\beta$ is the intrinsic peak width in radians ($2\theta$), $\theta$ is Bragg angle, and 0.88 is the Scherrer constant.[30]

The nanoparticles were imaged using a Philips CM-12 TEM at 100 keV with a $SiO_2$ grid. The grid was dipped in the Fe/Au nanoparticle saturated propanol solution, and the grid was dried in air, then in the oven at 130°C for 2 hours. Elemental analysis was performed by EDS which is attached to CM-12.

Additionally, the Fe@Au nanoparticles were studied by atomic-resolution Z-contrast imaging and electron energy loss spectroscopy (EELS) in a scanning transmission electron microscope (STEM). The STEM experiments were performed in a FEI Tecnai G2 Schottky field emission STEM/TEM operated at 200 KV and equipped with a post-column high resolution Gatan energy filtering (GIF) spectrometer which is located at the National Center for Electron Microscopy (NCEM) in the Lawrence Berkeley National Laboratory (LBNL). The optical conditions of the microscope for imaging and spectroscopy were defined to obtain a probe-size of 0.14 nm, with a convergence semi-angle of 13.5 mrad and a collection semi-angle of 136 mrad. In this experimental setup, the high-angle annular dark field (HAADF) image is predominantly incoherent and the image intensity is interpreted to be proportional to the atomic number square, $Z^2$.[31, 32] This condition, known as Z-contrast imaging, allows the structure and composition of the nanoparticles to be directly observed on the image and can also be used to position the electron probe for EELS.[33] Core loss EELS map the unoccupied density of states near the conduction band and is completely analogous to (XANES),[34] but with a much higher spatial resolution and it is only limited by the electron probe size.

Elemental analyses of Fe/Au nanoparticles were performed by ICP analysis of Fe and Au by Desert Analytics Laboratory in Tucson, Arizona. The sample was sent under nitrogen-filled and sealed vial.

*Magnetic and Transport Measurements*

Magnetic measurements were performed using a Quantum Design Superconducting QUantum Interference Device (SQUID) magnetometer, right after the synthesis. Approximately 40 mg of sample was placed in a gel-capsule, packed with glass wool and suspended in a straw. To prevent oxidation, the sample was immersed in degassed oil in the gel capsule under nitrogen.

For electrical transport measurements, pellets were prepared by cold-pressing nanoparticles into a 6 mm die under a 2 x$10^7$ Pa pressure for 10 minutes. Electrical leads were attached by silver paint onto the pressed pellet. The temperature-dependence of resistance, magnetoresistance at 5K, and saturation magnetization were measured repeatedly over 2 months to monitor the time scale of iron oxidation.

**Results and Discussion**

*X-Ray Diffraction*

As shown in Fig. 1, the patterns confirm the presence of both α-Fe and Au, with some of the peaks overlapping, consistent with previous reports.[17, 18] To investigate whether or not there is amorphous Fe or Fe-oxide present in the sample, the Fe/Au product was heated in air to 400°C and left at that temperature in air overnight. Any amorphous Fe will oxidize and crystallize and any Fe oxide present should become crystalline and be detectable by XRD. However, the diffraction pattern is quite similar to the original pattern (Fig. 1) and no new diffraction peaks are observed. This suggests that any Fe in the sample is coated in Au and that there is no amorphous Fe containing oxides as a by-product. It is also possible that any oxidized product is coated with Au and perhaps amorphous. The crystallite size of nanoparticle, calculated from the (111) Au reflection using the Scherrer formula[30] and calibrated for instrumentation width, is 19nm.

*Electron Micrographs*

Figure 1 inset shows a typical TEM image of Fe/Au nanoparticles. EDS confirm the presence of Fe and Au. As the nanoparticles are still magnetic at room temperature, they tend to aggregate on the grid and the image is blurred due to the interaction of the particles with the electron beam. The diameter and size distribution of final core/shell nanoparticles was measured by Analysis Soft imaging system to be 18±4 nm, consistent with the average size determined from peak broadening of the XRD pattern.

Figure 2 shows a high resolution Z-contrast image of a typical Fe/Au nanoparticle. Most of the nanoparticles show a darker region (lower contrast) usually located at the center of the nanoparticles. The pronounced difference of contrast shown within the nanoparticles by Z-contrast imaging indicates the difference in chemical composition within the nanoparticles. This difference of contrast is clear in Fig. 2 between the center of the nanoparticle and its edges. Au, as a heavy element, scatters electrons more strongly than Fe, which has a smaller atomic number. Consequently, in the Z-contrast image shown in Fig. 2, the brighter regions within the nanoparticle are Au rich while the darker regions are Fe rich and Au poor. Change of contrast can also be produced by change of thickness within the nanoparticle. However, EELS spectra taken on the two different regions do not show change in the background signal, indicating that the thickness is constant within the nanoparticle. Thus, this change of contrast is a strong indication that the nanoparticle is composed of a core Fe phase coated by Au. Nevertheless, the Z-contrast image alone cannot distinguish whether or not the core of the nanoparticle is metallic Fe or an oxide Fe phase.

As can be seen in the STEM image shown in Fig. 2, the Au coating is continuous, but exhibits topographical roughness on the nanometer scale. It can be hypothesized that the Au-shell grows by nucleating from small nanoparticles on the Fe-core surface before it develops the shell structure. In a report by Pham *et al.*,[35] chemical directing groups are placed on the surface of a silica nanoparticle and act as attachment points for small colloidal Au particles on the silica. They have shown that these nanoparticle nucleation sites form islands for the growth and coalescence of the thin Au overlayer. Here, we propose a similar mechanism without the addition of chemical directing agents for these Fe/Au core/shell nanoparticles. Specifically, $Au^{3+}$ is reduced to Au by $NaBH_4$, which initiates minimum nanoscaled seed Au nanoparticles and they grow larger resulting in an Au shell. The small colloidal particles of Au attach to the Fe core and template the growth of an Au overlayer. The rough surface may compromise the oxidation-resistance of the Au shell.

To further investigate the chemical composition of the nanoparticles, atomic-resolution EEL spectra were acquired. Figure 2a shows the O K-edge and Fe $L_{23}$-edge spectra from core and edge of a nanoparticle, as well as a spectrum from the silica film support (shown as reference only). Each spectrum is the sum of 8 individual spectra with an acquisition time of 10 seconds and an energy resolution of 3 eV. An energy dispersion of 1 eV/pixel was used. The spectra are summed up to increase the signal-to-noise ratio and background subtracted before the O K-edge onset. The O K-edge onset for all three spectra was determined to be at 532±1 eV.

Fe signal is present as a strong signal in the core spectrum as shown in Fig. 2a. The spectrum of the edge of the particle shows only a trace signal for Fe, however the signal is slightly above the noise level. The Fe signal at the edge of the nanoparticle is presumably coming from a residual Fe oxide phase around the Fe/Au nanoparticle as a result of the synthesis process. This may arise from inadequate rinsing of the nanoparticle or be due to Fe that does not get coated with Au that slowly oxidizes over time. This signal is low enough so as not to change the results of the analysis described below. To characterize the Fe oxidation state of the core, the $L_3/L_2$ white-line ratio was calculated. White-lines arise mainly from dipole selection rules due to transitions from the inner shell electrons to unoccupied states in the valence band.[34] The $L_3$ and $L_2$ white-lines or peaks result from transitions $2p^{3/2} \rightarrow 3d^{3/2}3d^{5/2}$ and $2p^{1/2} \rightarrow 3d^{3/2}$, respectively. The $L_3/L_2$ ratio was measured by the second derivative method, which has proven to characterize effectively Fe oxidation states.[36] The maximum of the two peaks on Fe core spectra are located at 709 eV and 722 eV, for $L_3$ and $L_2$, respectively. The $L_3/L_2$ ratio of the Fe core measured was 3.3±0.8. This value was compared to a set of reference data of $L_3/L_2$ ratios taken from specimens with known Fe oxidation states. Colliex *et. al.*[37] report for FeO, $Fe_3O_4$ and $\alpha$-$Fe_2O_3$ $L_3/L_2$ ratios of 3.9±0.8, 4.2±0.3 and 4.7±0.3, respectively. The $L_3/L_2$ ratio calculated for the Fe core nanoparticle is clear smaller than the Fe oxide phases reported by Colliex *et. al*. As a consequence, the Fe core nanoparticle is composed of a Fe metallic phase. Oxygen signal was found in all three spectra as it is shown in Fig. 2a. The O K-edge obtained from the film, which comes mainly from oxygen on the silica support, presents two main peaks, with maximum intensities at 536 eV and 560 eV, respectively. The spectra acquired from the core and edge of the nanoparticle also present these two peaks on the O K-edge, but with some differences. The core of the nanoparticle has a wider first peak than the edge or the silica support due to the increase of intensity of a post shoulder at 541 eV. Nevertheless, none of the O K-edge spectra obtained the nanoparticle have the features of any of the Fe oxide phases reported by Colliex *et. al.* indicating again that the core is formed by a metallic Fe phase. For instance, FeO, which has the closest features to the Fe core spectra as well as its $L_3/L_2$ ratio, presents a well defined pre-peak on the O K-edge which none of the nanoparticle spectra have.

*Magnetic Properties and Size Determination*

Magnetic hysteresis loops of the Fe/Au nanoparticles at 300 K and 5 K are shown in Fig. 3a. At 5K, the particles display a coercivity of 400 Oe, remanent magnetization of 14 emu/g, and a saturation magnetization $M_S$ of 43 emu/g. Correcting for the composition of the nanoparticles, 26.5 at.% of Fe as determined from inductively coupled plasma (ICP) analysis, the saturation magnetization is 162 emu/(g-Fe), close to the expected saturation moment of 220 emu/g for bulk Fe. At 300 K, the Fe/Au nanoparticles still exhibit significant saturation moment, about 2/3 of the 5 K $M_S$, although the hysteresis has diminished. These results suggest that we have some Fe cores that are large enough to behave like bulk Fe at room temperature.

Temperature dependence of the magnetization, after zero-field cooling (ZFC) and field cooling (FC), was measured in a 100 Oe field, as shown in Fig. 3b. Unlike earlier samples with smaller Fe cores which have low blocking temperatures, the present sample does not show clear blocking behavior up to 300 K. This indicates that the average magnetic core size is larger than the nanoparticles that were previously reported as 12 nm diameter with a blocking temperature of 150 K.[18] When the nanoparticles are well separated, they can be approximated by independent single domain particles and their thermally assisted magnetization reversal process can be described by $1/\tau = f_o e^{-KV/k_B T}$, where K is the magnetic anisotropy constant (Fe: $5 \times 10^5$ erg/cm$^3$), $f_o$ is a frequency factor ($10^9$ /s) and $\tau$ is the relaxation time (SQUID: ~30 s).[38] A 300 K blocking temperature should correspond to a mean magnetic core size of about 16 nm. However, in our measurement geometry, the particle-particle interaction cannot be neglected, due to the close proximity of the nanoparticles, contributing to a higher blocking temperature than expected for non-interacting particles. The absence of a clear blocking behavior thus could be due to both particle aggregation and a larger average Fe core size.

To clarify the time scale of oxidation of Fe/Au nanoparticles, the saturation magnetization $M_S$ was measured everyday for 5 days since right after synthesis. The nanoparticles were directly exposed to air, stored and measured in gel capsules during this study. After 5 days, $M_S$ has decreased to 50 % of its initial value right after synthesis (Fig. 3c & 3d).

*Electrical Transport*

In the previous study,[18] we have found that if the nanoparticles were left in air, in loose powder form, they oxidize over time. This was determined by measuring the electrical transport of pressed pellets made from these nanoparticles prepared immediately after synthesis and again one month later (pellet pressed from exposed particles). In this study, we first press the pellet and then keep the pellet in air and measure the electrical transport properties over time to monitor the stability of nanoparticles in the pressed pellet form.

Temperature dependence of the resistance of a pellet is shown in Fig. 4a. The resistance decreases slightly with decreasing temperature. This positive temperature coefficient of resistance is a signature of metallic conduction, in contrast to the negative temperature coefficient and thermally activated behavior seen in pellets of Fe-oxide nanoparticles.[39, 40] Furthermore, magnetoresistance (MR) has been measured at 5 K, as shown in Fig. 4b. Similar to our earlier study, a negative giant MR effect was observed, confirming the presence of magnetic scattering centers. These electrical measurements have been repeated many times over an 8-week period. The results obtained are always the same as those obtained right after synthesis. We note that the resistivity measurement is susceptible to a percolated conduction path through Au, thus less sensitive to Fe oxidation. In contrast, the MR effect is sensitive to Fe oxidation as it is due to spin-dependent scattering at the interface between Au and Fe as well as within the magnetic Fe core. Any oxidation of the Fe core, into magnetic or non-magnetic Fe-oxides, will change this spin-dependent scattering process and result in a change in MR. The lack of appreciable changes in both resistivity and MR results demonstrates that when pressed into a pellet, although still exposed to air, the Fe/Au nanoparticles are stable over time.

**Conclusion**

In summary, we have synthesized Fe-core/Au-shell nanoparticles by a reverse micelle method, and investigated their growth mechanisms and oxidation-resistant characteristics. The core/shell heterostructure and the presence of the Fe & Au phases have been clearly confirmed. The Au shell appears to grow by nucleating at selected sites on the Fe core surface before coalescing. The rough surface could compromise the oxidation-resistance of the Au shell. Indeed, the magnetic moments of such nanoparticles, in the loose powder form, decrease over time due to oxidation. The oxidized product does not show crystalline Fe oxides in the powder diffraction pattern. In the pressed pellet form, electrical transport measurements show that the particles are fairly stable, as the resistance and magnetoresistance of the pellet do not change appreciably over time. These results provide direction for new synthesis routes to achieve truly airtight Au-shells over Fe-cores.

## Acknowledgements

This work was supported by the National Science Foundation (DMR-0120990 and CHE-0210807), American Chemical Society (PRF-39153-G5B), University of California (CLE), the Ministère de la Région Wallonne (115012), the U.S. Department of Energy, Division of Materials Sciences, Office of Basic Energy Science, under Contract No. FG02 03ER46057. J. O. is supported by a NEAT-IGERT fellowship. The FEI Tecnai F20 located at National Center for Electron Microscopy at LBNL is supported by LDRD funding from LBNL. The authors thank Peter Klavins (Physics, UC Davis) for his technical contribution and helpful discussions.

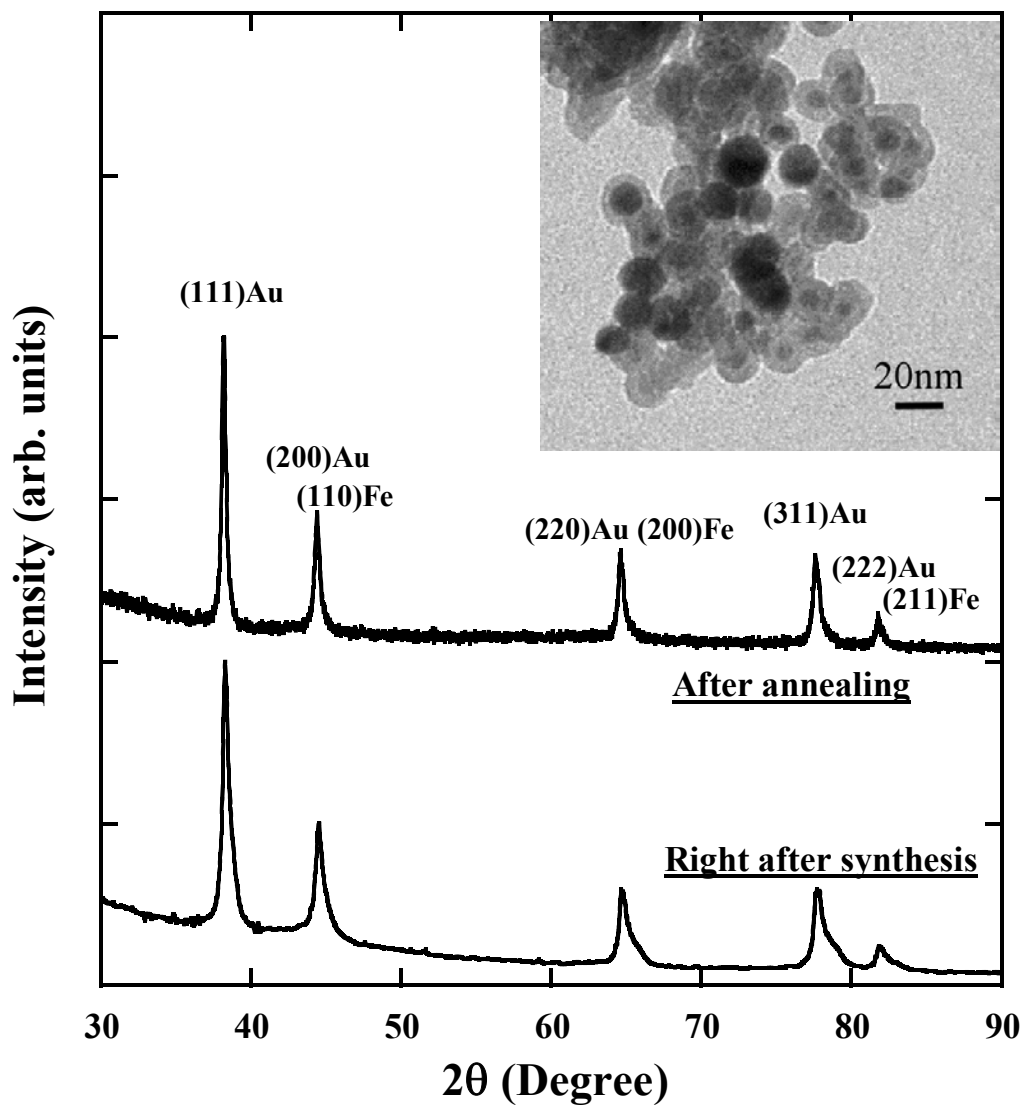

**Figure 1.** X-ray diffraction patterns of Au-coated nanoparticles, right after synthesis and after overnight annealing at 400°C in air. The inset shows a transmission electron microscopy image of Au-coated Fe nanoparticles.

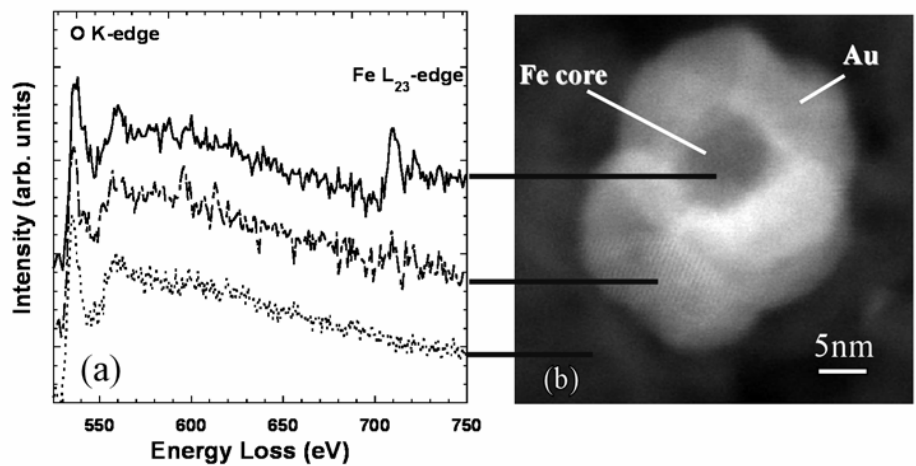

**Figure 2.** High resolution Z-Contrast image of an Au-coated Fe nanoparticle obtained by scanning transmission electron microscopy, the corresponding Oxygen K-edge and Fe $L_{23}$-edge spectra acquired from the center (solid) and surface (dashed) of the Fe/Au nanoparticle, and the silica film support (dotted). The nanoparticle core is composed predominantly of a Fe metallic phase.

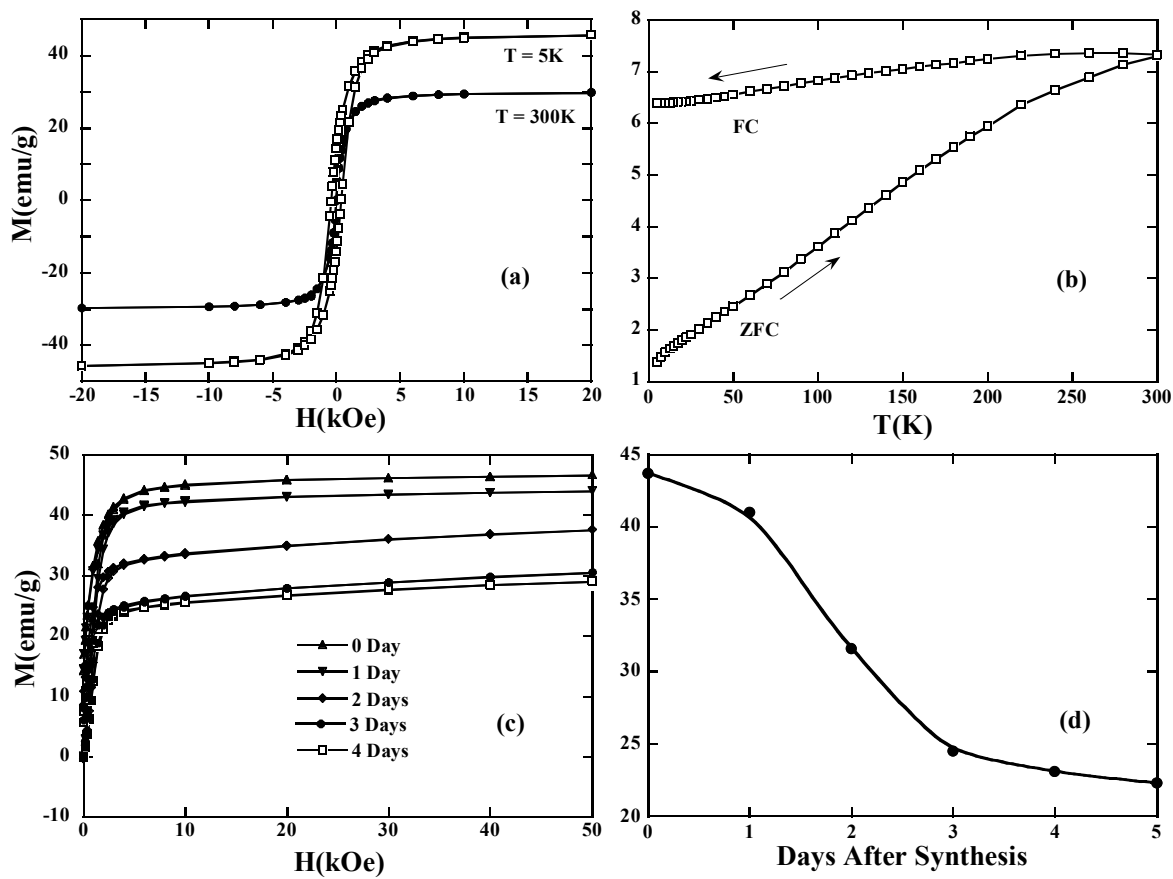

**Figure 3.** (a) Magnetic hysteresis loop at 5K and 300K. (b) Temperature dependence of the magnetization, after zero field cooling (ZFC) and field cooling (FC), measured in a 100 Oe field. (c) First quandrant of the magenetic hysteresis loop at 5 K. From the top, each curve indicates the measurement with 1 day interval right after synthesis. (d) Decay of saturation magnetization of exposed Fe/Au nanoparticles over time.

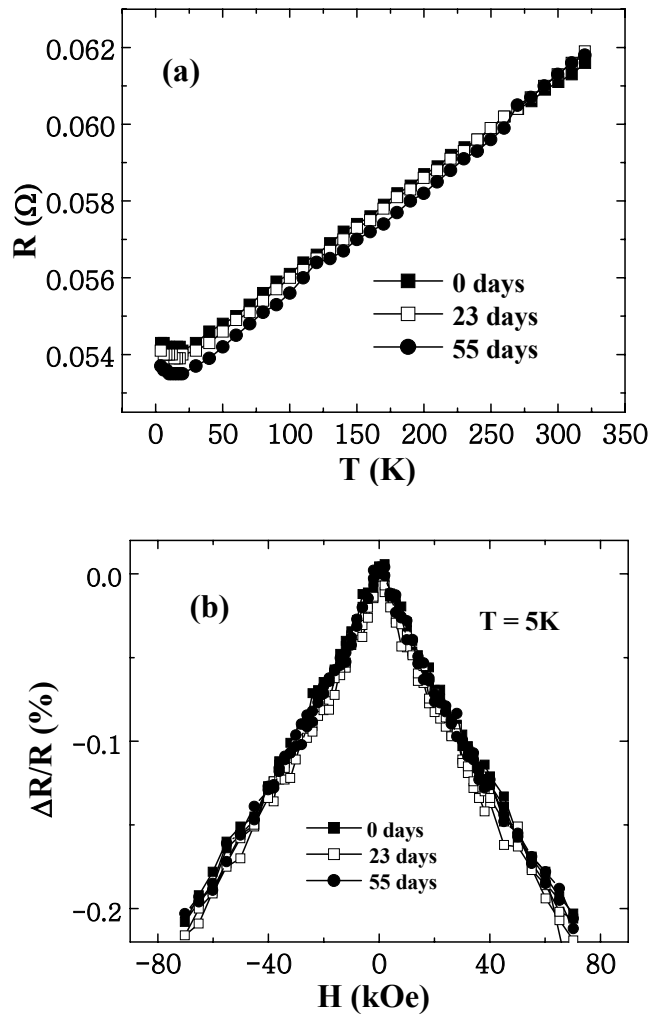

**Figure 4.** (a) Temperature dependence of resistance in zero magnetic field and (b) field dependence of magnetoresistance at 5 K of a pressed pellet of Au-coated Fe nanoparticles, measured at different times after synthesis.

21